# NIST 和 IEEE 金鑰衍生函數標準的比較和實作

Abel C. H. Chen
*Information & Communications Security Laboratory,*
*Chunghwa Telecom Laboratories*
Taoyuan, Taiwan
ORCID: 0000-0003-3628-3033

*摘要*—由於許多應用和服務可能都會用到偽隨機數，在給定特定金鑰值和特定訊息的條件下，可以通過金鑰衍生函數(Key Derivation Function, KDF)產製特定的偽隨機數(Pseudorandom Number, PRN)。其中，金鑰衍生函數主要建構在訊息驗證碼(Message Authentication Code)的基礎上，通過訊息驗證碼來組成偽隨機數。有鑑於此，本研究首先討論美國國家標準暨技術研究院(National Institute of Standards and Technology, NIST)定義的基於雜湊訊息驗證碼(Keyed-Hash Message Authentication Code, HMAC)、基於密文訊息驗證碼(Cipher-based Message Authentication Code, CMAC)、以及基於 Keccak 訊息驗證碼(Keccak-based Message Authentication Code, KMAC)。之後再討論基於前述訊息驗證碼的金鑰衍生函數，包含基於計數器模式(Counter Mode)金鑰衍生函數、基於 KMAC 金鑰衍生函數、以及 IEEE 1609.2.1 定義的金鑰衍生函數。在實驗中，本研究比較各種訊息驗證碼產製的計算時間和各種基於金鑰衍生函數的偽隨機數產製的計算時間，再討論每個方法的優缺點及其適用場域。其中，由實驗結果顯示，基於密文訊息驗證碼和基於 CMAC 金鑰衍生函數的計算時間最短，平均僅需約 0.007 毫秒和 0.014 毫秒。

*關鍵字*—金鑰衍生函數、訊息驗證碼、基於雜湊、基於密文、基於*Keccak*

## I. 前言

偽隨機數是許多資訊安全服務和應用的基礎功能，如何在特定特定金鑰值和特定訊息的條件下，可以通過金鑰衍生函數(Key Derivation Function, KDF)產製特定且安全的偽隨機數(Pseudorandom Number, PRN)是一個重要的議題[1]。其中，金鑰衍生函數主要建構在訊息驗證碼(Message Authentication Code)的基礎上，通過訊息驗證碼來組成偽隨機數[2]。並且，近幾年量子計算技術日益成熟，設計一個具備抗量子計算攻擊的訊息驗證碼和金鑰衍生函數將是主要挑戰之一[3]。

由於目前量子計算尚無法快速破解雜湊(Hash)演算法和進階加密標準(Advanced Encryption Standard, AES)演算法[4]，所以美國國家標準暨技術研究院(National Institute of Standards and Technology, NIST)在雜湊演算法和進階加密標準演算法基礎上分別設計基於雜湊訊息驗證碼(Keyed-Hash Message Authentication Code, HMAC) [5]-[6]、基於密文訊息驗證碼(Cipher-based Message Authentication Code, CMAC) [7]-[8]、以及基於 Keccak 訊息驗證碼(Keccak-based Message Authentication Code, KMAC) [9]-[10]。並且再於前述訊息驗證碼基礎上設計基於計數器模式金鑰衍生函數[11]和基於 KMAC 金鑰衍生函數[12]，提供具備抗量子計算攻擊的訊息驗證碼和金鑰衍生函數[13]。另外，在車聯網通訊環境，IEEE 1609.2.1 標準中定義從毛蟲金鑰擴展為繭金鑰時，可以讓終端設備和註冊中心具有共同的金鑰衍生函數來產製偽隨機數作為擴展值[14]，並且該金鑰衍生函數主要建構在進階加密標準演算法基礎上來保障其安全性[15]。

有鑑於此，本研究主要從原理和實作上來討論各種訊息驗證碼的金鑰衍生函數的流程和優缺點。本研究的主要貢獻條列如下：

- 本研究討論和實作了美國國家標準暨技術研究院定義的三種訊息驗證碼，包含基於雜湊訊息驗證碼[5]、基於密文訊息驗證碼[7]、以及基於 Keccak 訊息驗證碼[9]。

- 本研究討論和實作了三種金鑰衍生函數，包含美國國家標準暨技術研究院定義的基於計數器模式金鑰衍生函數和基於 KMAC 金鑰衍生函數[13]、以及 IEEE 1609.2.1 定義的金鑰衍生函數[15]。

- 本研究比較上述各種訊息驗證碼和各種金鑰衍生函數的計算時間，並且討論優缺點和適用情境。

本文主要分為五節。第 II 節介紹各種訊息驗證碼的流程和原理，並且第 III 節介紹各種金鑰衍生函數的流程和原理。第 IV 節描述實驗環境和討論實驗結果。最後，第 V 節總結本研究發現，並且討論未來研究方向。

## II. 訊息驗證碼

本節將分別對基於雜湊訊息驗證碼、基於密文訊息驗證碼、以及基於 Keccak 訊息驗證碼展開描述。其中，本節採用的金鑰值為 $k$、訊息值為 $Msg$，各小節皆以此參數進行討論。

### A. 基於雜湊訊息驗證碼

美國國家標準暨技術研究院提出的基於雜湊訊息驗證碼定義在文件 FIPS 198-1 [5]。基於雜湊訊息驗證碼產製流程如圖 1 所示。其中，$Hash(\cdot)$ 函數表示為安全雜湊演算法(Secure Hash Algorithm, SHA)，例如：SHA-256。而 $RightPad(\cdot)$ 函數表示為填充函數，對該輸入值右邊填充 0x00 直到達到區塊(block)長度 $B$。首先，為了讓金鑰長度(即 $Length(k)$)可以符合安全雜湊演算法區塊(block)長度 $B$ (例如：在 SHA-256 中 $B$ 為 32 bytes)運用公式(1)把金鑰值 $k$ 轉換為金鑰值 $K_0(k)$。之後分別運用公式(2)和公式(3)計算內部值 $Inner(k, Msg)$ 和外部值 $Outer(k)$；其中，$Ipad$ 值每個 byte 值是 0x36 並且重覆 $B$ 次，$Opad$ 值每個 byte 值是 0x5C 並且重覆 $B$ 次。最後，再根據內部值和外部值輸入到安全雜湊演算法計算後得到 $Hmac(k, Msg)$，如公式(4)所示。

$$K_0(k) = \begin{cases} RightPad(Hash(k)), \text{if } Length(k) > B \\ k, \text{if } Length(k) = B \\ RightPad(k), \text{if } Length(k) < B \end{cases} \quad (1)$$

$$Inner(k, Msg) = Hash\big((K_0(k) \oplus Ipad) || Msg\big) \quad (2)$$

$$Outer(k) = K_0(k) \oplus Opad \tag{3}$$

$$Hmac(k, Msg) = Hash\big(Outer(k)\|Inner(k, Msg)\big) \tag{4}$$

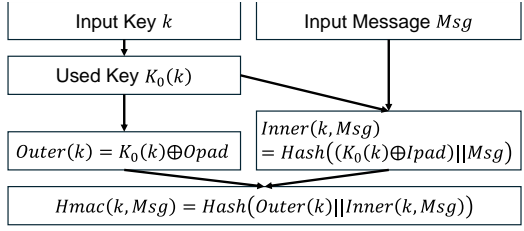

Fig. 1. The generation of hash-based message authentication code.

*B. 基於密文訊息驗證碼*

美國國家標準暨技術研究院提出的基於密文訊息驗證碼定義在文件 NIST Special Publication (SP) 800-38B [7]。基於密文訊息驗證碼產製流程如圖 2 所示。其中，$AES(key, plaintext)$函數表示為進階加密標準演算法，例如：AES-128。首先，分別運用公式(5)和公式(6)，執行進階加密標準演算法代表金鑰值$k$和明文值0，得到的密文值往左移位 1 個位元後得到$K_1(k)$，並且對$K_1(k)$往左移位 1 個位元後得到$K_2(k)$。之後根據區塊長度$B$，切訊息$Msg$為子訊息集合$\{M_1, M_2, \dots, M_m\}$，讓每個子訊息長度符合區塊長度，如公式(7)所示。其中，運用公式(8)讓最後一個子訊息長度也能符合區塊長度，並且與$K_1(k)$或$K_2(k)$作用來混淆。之後執行密文區塊鏈的方式加密每一個子訊息，並且把前一個區塊密文$c_i$與當下區塊明文$M_{i+1}$做邏輯互斥或(exclusive or, XOR)計算後，再用金鑰值 $k$ 運作加密標準演算法加密，如公式(9)所示。把$m$個子訊息都做過計算後可得$Cmac(k, Split(Msg))$，如公式(10)所示。

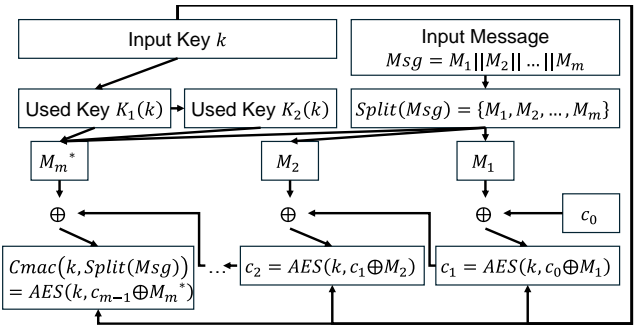

Fig. 2. The generation of cipher-based message authentication code.

$$K_1(k) = AES\big(k, RightPad(0)\big) \ll 1 \tag{5}$$

$$K_2(k) = K_1(k) \ll 1 \tag{6}$$

$$Split(Msg) = \{M_1, M_2, \dots, M_m\},$$
$$\text{where } m = \left\lceil \frac{Length(Msg)}{B} \right\rceil \tag{7}$$

$$M_m^* = \begin{cases} M_m \oplus K_1(k), & \text{if } Length(M_m) = B \\ RightPad1(M_m) \oplus K_2(k), & \text{if } Length(M_m) < B \end{cases} \tag{8}$$

$$c_{i+1} = AES(k, c_i \oplus M_{i+1}),$$
$$\text{where } c_0 = RightPad(0) \text{ and } 0 \le i < m \tag{9}$$

$$Cmac(k, Split(Msg)) = AES(k, c_{m-1} \oplus M_m^*) \tag{10}$$

*C. 基於Keccak 訊息驗證碼*

美國國家標準暨技術研究院提出的基於 Keccak 訊息驗證碼定義在文件 NIST SP 800-185 [9]。基於 Keccak 訊息驗證碼產製流程如圖 3 所示，主要建構在可客製化安全雜湊演算法 Keccak (customizable Secure Hash Algorithm Keccak, cSHAKE)方法，而可客製化安全雜湊演算法 Keccak 方法則是修改自安全雜湊演算法 Keccak (Secure Hash Algorithm Keccak, SHAKE) [16]。首先，將先採用公式(11)把金鑰值$k$和訊息值$Msg$轉換為$Msg^*$；其中，$L$表示為採用的位元長度、$RightPad(k, r)$表示為對k值右邊填充 0x00 直到達到長度$r$。然後再執行公式(12)，把$Msg^*$作為待被雜湊計算的訊息代入$CSHAKE(Msg^*, L, "KMAC", S)$ 函數；其中，$CSHAKE(Msg^*, L, "KMAC", S)$函數表示為 cSHAKE 方法、在基於 Keccak 訊息驗證碼應用代入名稱為"KMAC"的字串 byte 陣列、以及$S$表示為可客製化的字串 byte 陣列(例如：在基於 KMAC 金鑰衍生函數應用$S = "KDF"$)。

$$Msg^* = RightPad(k, r)\|Msg\|RightEnc(L),$$
$$\text{where } L \in \{256, 512\} \text{ and } r = \begin{cases} 168, & \text{if } L = 256 \\ 136, & \text{if } L = 512 \end{cases} \tag{11}$$

$$Kmac(k, Msg, L, S) = CSHAKE(Msg^*, L, "KMAC", S) \tag{12}$$

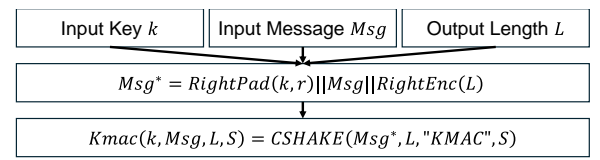

Fig. 3. The generation of Keccak-based message authentication code.

III. 金鑰衍生函數

本節將分別介紹 NIST 定義的基於計數器模式金鑰衍生函數和基於 KMAC 金鑰衍生函數，以及 IEEE 1609.2.1 定義的基於計數器金鑰衍生函數。本節採用的金鑰值為$k$、訊息值為 $Msg$、輸出值長度為$L$，各小節皆以此參數進行討論。

*A. NIST 定義的基於計數器模式金鑰衍生函數*

美國國家標準暨技術研究院提出的基於計數器模式金鑰衍生函數定義在文件 NIST SP 800-108r1-upd1 [13]。基於計數器模式金鑰衍生函數的偽隨機數產製流程如圖 4 所示。其中，$PRF(\cdot)$函數表示為偽隨機函數(Pseudorandom Function, PRF)，偽隨機函數可以是基於

雜湊訊息驗證碼或基於密文訊息驗證碼。計數器的運作上主要將根據輸出值長度為$L$和訊息驗證碼區塊長度$B$產製$n$個訊息驗證碼，最後的輸出值再把每個訊息驗證碼連結(concatenate)起來，如公式(13)所示。其中，第$i$個訊息驗證碼的輸入訊息值為$i||"KDF"||0x00||Msg||L$。

$$\begin{aligned}&Ckdf(k,Msg,L)\\&=PRF(k,Msg'_1)||\ldots||PRF(k,Msg'_n),\text{ where}\\&n=\left\lceil\frac{L}{B}\right\rceil,\\&Msg'_i=i||"KDF"||0x00||Msg||L,\\&\text{and }PRF(k,Msg)\\&\in\{Hmac(k,Msg),Cmac(k,Split(Msg))\}\end{aligned} \quad (13)$$

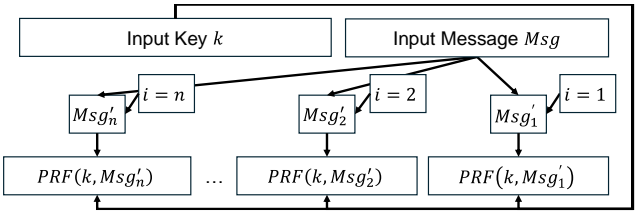

Fig. 4. The procedure of key derivation function in counter mode defined by NIST.

為方便說明，偽隨機函數是基於雜湊訊息驗證碼的基於計數器模式金鑰衍生函數，本研究命名其為"基於雜湊訊息驗證碼金鑰衍生函數(HMAC-based KDF)"。偽隨機函數是基於密文訊息驗證碼的基於計數器模式金鑰衍生函數，本研究命名其為"基於密文訊息驗證碼金鑰衍生函數(CMAC-based KDF)"。值得注意的是，基於雜湊訊息驗證碼金鑰衍生函數使用的雜湊演算法可以是第二代安全的雜湊演算法(SHA-2)，也可以是第三代安全的雜湊演算法(SHA3)，例如：SHA-256和SHA3-256。

*B. NIST 定義的基於 KMAC 金鑰衍生函數*

美國國家標準暨技術研究院提出的基於 KMAC 金鑰衍生函數定義在文件 NIST SP 800-108r1-upd1 [13]。基於 KMAC 金鑰衍生函數的偽隨機數產製流程如圖 5 所示。可以發現其主要跟基於 Keccak 訊息驗證碼產製流程一致，差異在於設定$S="KDF"$。

$$\begin{aligned}&Kkdf(k,Msg,L)\\&=Kmac(k,Msg,L,"KDF")\\&=CSHAKE(Msg^*,L,"KMAC","KDF")\end{aligned} \quad (14)$$

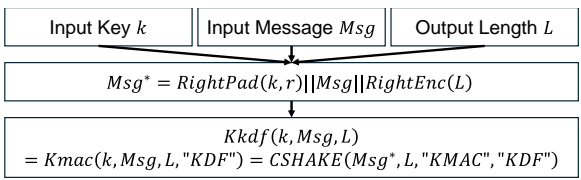

Fig. 5. The procedure of key derivation function using KMAC defined by NIST.

*C. IEEE 1609.2.1 定義的基於計數器金鑰衍生函數*

IEEE 為了在車聯網通訊環境提供終端設備匿名性，所以定義了另一套基於計數器金鑰衍生函數[15]。IEEE 1609.2.1 定義的基於計數器金鑰衍生函數的偽隨機數產製流程如圖 6 所示。讓終端設備可以分享 AES-128 金鑰值$k$給註冊中心，並且訊息 $Msg$ 為雙方共同可知的 $iValue$ 和 $jValue$；其中，$iValue$ 是 4-byte 長度的週期索引值、$jValue$ 是 4-byte 長度的金鑰索引值。在數位簽章用途上設定$U=1$，而在加解密用途上設定$U=2$。並且，當$U=1$，執行$RightPad(0)$產生 4 個 byte 的 0x00；當$U=2$，執行$RightPad2(1)$產生 4 個 byte 的 0x11。之後分別計算 $i$ 從 1 到 3 的 $Msg''_i$，分別產生訊息驗證碼 $(AES(k,Msg''_i)\oplus Msg''_i)$，最後再把 3 個訊息驗證碼連結(concatenate)起來，如公式(15)所示。通過 IEEE 1609.2.1 定義的基於計數器金鑰衍生函數運作後，終端設備可以得到偽隨機數擴展其毛蟲私鑰為繭私鑰，而註冊中心則可以用相同的偽隨機數擴展該終端設備的毛蟲公鑰為繭公鑰，並且繭金鑰和繭私鑰成對。

$$\begin{aligned}&Ikdf(k,Msg,U)=(AES(k,Msg''_1)\oplus Msg''_1)||\\&\qquad\qquad\qquad\quad(AES(k,Msg''_2)\oplus Msg''_2)||\\&\qquad\qquad\qquad\quad(AES(k,Msg''_3)\oplus Msg''_3),\\&\text{where }Pad_U=\begin{cases}RightPad(0),\text{ if }U=1\\RightPad2(1),\text{ if }U=2\end{cases}\\&\text{and }Msg''_i=(Pad_U||Msg||RightPad(0))+i\end{aligned} \quad (15)$$

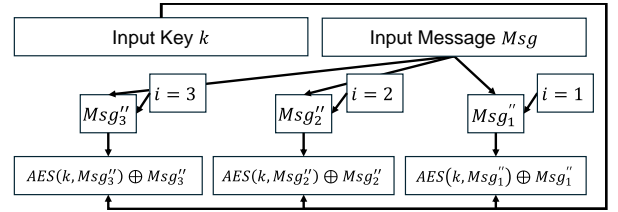

Fig. 6. The procedure of key derivation function in counter mode defined by IEEE.

IV. 實驗結果與討論

第 IV.A 節將先介紹實驗環境，之後分別在第 IV.B 節和第 IV.C 節討論訊息驗證碼產製和基於金鑰衍生函數的偽隨機數產製的實驗結果及其比較討論。

*A. 實驗環境*

為驗證各種訊息驗證碼和各種金鑰衍生函數的計算效率，本研究分別實作該些方法，並且在下列的硬體和軟體的環境執行。

- CPU：Intel® Core™ i7-10510U
- RAM：16 GB
- Operating System：Windows 11 Enterprise Edition
- Java Environment：OpenJDK 21.0.1
- Cryptographic Library：BouncyCastle 1.70

## B. 訊息驗證碼產製實驗結果與比較

在訊息驗證碼產製實驗中，本研究主要比較三種訊息驗證碼，包含基於雜湊訊息驗證碼、基於密文訊息驗證碼、以及基於 Keccak 訊息驗證碼。並且，有鑑於 IEEE 1609.2.1是採用 AES-128，所以為公平比較基於雜湊訊息驗證碼採用 SHA-256、基於密文訊息驗證碼採用 AES-128、基於 Keccak 訊息驗證碼採用 cSHAKE-128。運用不同的方法產製 1000 個訊息驗證碼的計算時間平均值、中位數、以及標準差整理如表 I 所示，單位為毫秒；並且為了充分展示資料分佈，本研究將上述實驗結果以盒鬚圖型式表示於圖 7。由實驗結果可以觀察到基於密文訊息驗證碼的計算時間最短，具有最高的執行效率。然而，基於 Keccak 訊息驗證碼則大約需要基於密文訊息驗證碼兩倍的計算時間。

TABLE I.　THE COMPUTATION TIME COMPARISON OF MESSAGE AUTHERTICATION CODE GENERATION (UNIT: MILLISECONDS)

|  | HMAC | CMAC | KMAC |
|---|---|---|---|
| Mean | 0.007 | 0.007 | 0.015 |
| Median | 0.006 | 0.006 | 0.014 |
| Standard Deviation | 0.005 | 0.005 | 0.008 |

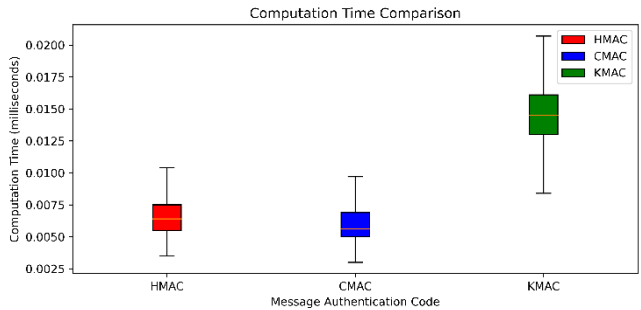

Fig. 7.　The computation time comparison of message autherntication code generation.

## C. 基於金鑰衍生函數的偽隨機數產製實驗結果與比較

在基於金鑰衍生函數的偽隨機數產製實驗中，本研究主要比較四種金鑰衍生函數，包含基於雜湊訊息驗證碼金鑰衍生函數、基於密文訊息驗證碼金鑰衍生函數、基於 KMAC 金鑰衍生函數、以及 IEEE 1609.2.1 定義的金鑰衍生函數。由於 IEEE 1609.2.1 是採用 AES-128，所以基於雜湊訊息驗證碼金鑰衍生函數採用 SHA-256、基於密文訊息驗證碼金鑰衍生函數採用 AES-128、基於 KMAC 金鑰衍生函數採用 cSHAKE-128。運用不同的方法產製 1000 個 48-byte 長度偽隨機數的計算時間平均值、中位數、以及標準差整理如表 II 所示，單位為毫秒；並且為了充分展示資料分佈，本研究將上述實驗結果以盒鬚圖型式表示於圖 8。由實驗結果可以觀察到基於密文訊息驗證碼金鑰衍生函數的計算時間最短，具有最高的執行效率。其次分別為基於雜湊訊息驗證碼金鑰衍生函數和基於 KMAC 金鑰衍生函數。然而，IEEE 1609.2.1 定義的金鑰衍生函數需要最多的計算時間，其原因在於其除了用 AES 進行加密計算之外，還需要把結果與訊息值再做一次 XOR (詳見公式(15))，導致需要比其他方法更多的計算時間。

除此之外，IEEE 1609.2.1 定義的金鑰衍生函數是採用 AES 電子密碼本(Electronic Code Book, ECB)模式加密，將可能造成安全疑慮。並且，雖然 IEEE 1609.2.1 定義的金鑰衍生函數相較於美國國家標準暨技術研究院定義的金鑰衍生函數，多做了與訊息值 XOR 計算。但由於訊息值在應用情境中可以視為公開資訊，所以多做了 XOR 計算對安全性的提升有限。因此，未來可以考慮在車聯網通訊環境中結合美國國家標準暨技術研究院定義的金鑰衍生函數來提供鞼金鑰擴展函數。

TABLE II.　THE COMPUTATION TIME COMPARISON OF PSEUDORANDOM NUMBER GENERATION BASED ON KEY DERIVATION FUNCTIONS (UNIT: MILLISECONDS)

|  | HMAC-based KDF | CMAC-based KDF | KMAC-based KDF | KDF in IEEE 1609.2.1 |
|---|---|---|---|---|
| Mean | 0.021 | 0.014 | 0.038 | 0.069 |
| Median | 0.016 | 0.014 | 0.039 | 0.065 |
| Standard Deviation | 0.106 | 0.009 | 0.016 | 0.037 |

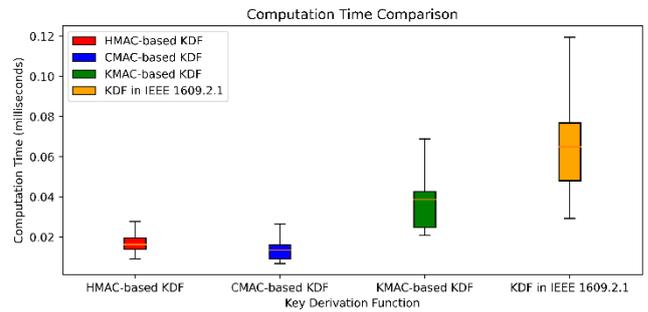

Fig. 8.　The computation time comparison of pseudorandom number generation based on key derivation functions.

## V.　結論與未來研究

本研究從原理和實作比較了三種訊息驗證碼，包含基於雜湊訊息驗證碼、基於密文訊息驗證碼、以及基於 Keccak 訊息驗證碼。同時在此基礎上，比較了四種金鑰衍生函數，包含基於雜湊訊息驗證碼金鑰衍生函數、基於密文訊息驗證碼金鑰衍生函數、基於 KMAC 金鑰衍生函數、以及 IEEE 1609.2.1 定義的金鑰衍生函數。由實驗結果發現基於密文訊息驗證碼的計算時間最短，而基於雜湊訊息驗證碼的計算時間次短。除此之外，在金鑰衍生函數的計算效率上，基於密文訊息驗證碼金鑰衍生函數計算最快，基於雜湊訊息驗證碼金鑰衍生函數次快，而 IEEE 1609.2.1 定義的金鑰衍生函數計算最慢。本研究已在第 IV.C 節討論了 IEEE 1609.2.1 定義的金鑰衍生函數的安全疑慮和限制。

在未來研究中，由於為達到車聯網通訊環境中的終端設備匿名性，金鑰衍生函數是必備的工具之一。因此，未來可以考慮改採用美國國家標準暨技術研究院定義的金鑰衍生函數，並且可以考慮是採用基於雜湊訊息驗證碼金鑰衍生函數，以兼顧計算效率又可以避免 AES 電子密碼本模式的安全疑慮。

參考文獻

[1]　J. M. Mcginthy and A. J. Michaels, "Further Analysis of PRNG-Based Key Derivation Functions," in *IEEE Access*, vol. 7, pp. 95978-95986, 2019, doi: 10.1109/ACCESS.2019.2928768.

[2]　H. Sikarwar and D. Das, "A Novel MAC-Based Authentication Scheme (NoMAS) for Internet of Vehicles (IoV)," in *IEEE Transactions on Intelligent Transportation Systems*, vol. 24, no. 5, pp. 4904-4916, May 2023, doi: 10.1109/TITS.2023.3242291.

[3] J. Jiang and D. Wang, "QPASE: Quantum-Resistant Password-Authenticated Searchable Encryption for Cloud Storage," in *IEEE Transactions on Information Forensics and Security*, vol. 19, pp. 4231-4246, 2024, doi: 10.1109/TIFS.2024.3372804.

[4] H. Sikarwar and D. Das, "Fast AES-Based Universal Hash Functions and MACs," in *IACR Transactions on Symmetric Cryptology*, vol. 2024, no. 2, pp. 35-67, June 2024, doi: 10.46586/tosc.v2024.i2.35-67.

[5] "The Keyed-Hash Message Authentication Code (HMAC)," in *FIPS 198-1*, pp.1-7, July 2008, doi: 10.6028/NIST.FIPS.198-1.

[6] H. Choi and S. C. Seo, "Optimization of PBKDF2 Using HMAC-SHA2 and HMAC-LSH Families in CPU Environment," in *IEEE Access*, vol. 9, pp. 40165-40177, 2021, doi: 10.1109/ACCESS.2021.3065082.

[7] M. Dworkin, "Recommendation for Block Cipher Modes of Operation: The CMAC Mode for Authentication," in *NIST Special Publication 800-38B*, pp. 1-16, 2016, doi: 10.6028/NIST.SP.800-38B.

[8] P. Nannipieri et al., "VLSI Design of Advanced-Features AES Cryptoprocessor in the Framework of the European Processor Initiative," in *IEEE Transactions on Very Large Scale Integration (VLSI) Systems*, vol. 30, no. 2, pp. 177-186, Feb. 2022, doi: 10.1109/TVLSI.2021.3129107.

[9] J. Kelsey, S. Chang, R. Perlner, "SHA-3 Derived Functions: cSHAKE, KMAC, TupleHash and ParallelHash," in *NIST Special Publication 800-185*, pp. 1-26, 2016, doi: 10.6028/NIST.SP.800-185.

[10] Z. Li et al., " New Conditional Cube Attack on Keccak Keyed Modes," in *IACR Transactions on Symmetric Cryptology*, vol. 2019, no. 2, pp. 94-124, June 2019, doi: 10.13154/tosc.v2019.i2.94-124.

[11] D. Song, Y. Yan, G. Shao, F. Zhu and M. Song, "A Security Analysis of a Deterministic Key Generation Scheme," *Proceedings of 2024 IEEE International Conference on Cyber Security and Resilience (CSR)*, London, United Kingdom, 2024, pp. 309-314, doi: 10.1109/CSR61664.2024.10679483.

[12] J. -H. Phoon, W. -K. Lee, D. C. . -K. Wong, W. -S. Yap, B. -M. Goi and R. C. . -W. Phan, "Optimized IoT Cryptoprocessor Based on QC-MPDC Key Encapsulation Mechanism," in *IEEE Internet of Things Journal*, vol. 7, no. 9, pp. 8513-8524, Sept. 2020, doi: 10.1109/JIOT.2020.2991334.

[13] L. Chen, "Recommendation for Key Derivation Using Pseudorandom Functions," in *NIST Special Publication 800-108r1-upd1*, pp. 1-26, 2016, doi: 10.6028/NIST.SP.800-108r1-upd1.

[14] A. C. H. Chen, "RSA-Based Anonymous Certificate for Security Credential Management System," *Proceedings of 2024 7th International Conference on Circuit Power and Computing Technologies (ICCPCT)*, Kollam, India, 2024, pp. 422-425, doi: 10.1109/ICCPCT61902.2024.10672686.

[15] "IEEE Standard for Wireless Access in Vehicular Environments (WAVE) - Certificate Management Interfaces for End Entities," in *IEEE Std 1609.2.1-2022 (Revision of IEEE Std 1609.2.1-2020)*, pp.1-261, 30 June 2022, doi: 10.1109/IEEESTD.2022.9810154.

[16] "SHA-3 Standard: Permutation-Based Hash and Extendable-Output Functions," in *FIPS 202*, pp.1-29, August 2015, doi: 10.6028/NIST.FIPS.202.